\documentclass[final,english]{bullsrsl}[2022/06/15]

\usepackage[latin1]{inputenc}
\usepackage[T1]{fontenc}

\usepackage{natbib} 
\usepackage{graphicx}
\usepackage{dingbat}
\usepackage[dvipsnames]{xcolor}
\usepackage{caption}
\usepackage{subcaption}
\newcommand{\sw}[1]{\texttt{#1}}

\begin{document}
\title{Insights into the properties of GRBs with TeV emission}

\author[affil={1}, corresponding]{Kuntal}{Misra}
\author[affil={1,2}]{Dimple}{}
\author[affil={1,3}]{Ankur}{Ghosh}

\affiliation[1]{Aryabhatta Research Institute of Observational Sciences (ARIES), Manora Peak, Nainital-263002, India.}
\affiliation[2]{Department of Physics, Deen Dayal Upadhyaya Gorakhpur University, Gorakhpur-273009, India.}
\affiliation[3]{School of Studies in Physics and Astrophysics, Pandit Ravishankar Shukla University, Raipur 492010, Chattisgarh, India}

\correspondance{kuntal@aries.res.in}
\date{13th October 2020}
\maketitle


%

\begin{abstract}
This study investigates the environments and characteristics of Gamma-Ray Bursts (GRBs) exhibiting very high energy (VHE) emission. Recent detections of VHE emission, up to TeV energies, challenge synchrotron-only emission models and particle acceleration concepts in GRBs. Until now, only a handful of GRBs have been detected in the VHE range. We compare the number densities of the circumburst medium of VHE-detected GRBs to check if the environment impacts the VHE emission. This shows that these GRBs have environments similar to the larger population of GRBs. We employ machine learning algorithms to create two-dimensional embeddings of GRB prompt emission light curves from the {\it Swift}-BAT catalog. VHE-detected GRBs are located across the map, indicating that VHE emission does not favour any particular cluster. These findings indicate that VHE-detected GRBs do not show any peculiar characteristics other than the observational detection of VHE photons. Future detections will increase the sample size required for a rigorous understanding of the origin of VHE emission in GRBs.

\end{abstract}

\keywords{GRBs, VHE emission, emission mechanisms, environments, Machine Learning}

\section{Introduction}
\label{introduction}

Gamma-Ray Bursts (GRBs) are brief, luminous flashes of gamma-rays arising from extreme cataclysmic events in the universe. Though GRBs have been studied extensively for decades, the emission mechanisms powering them remain uncertain. The relativistic fireball shock model is widely accepted, where a relativistic outflow dissipates its kinetic energy through internal or external shocks, producing GRB emission \citep{RevModPhys.76.1143, 2015PhR...561....1K}. But details of the microphysical dissipation and radiation processes are still debated. Suggested emission mechanisms invoke dissipation through internal or external shocks. Broadband spectra and variability of GRB prompt emission indicate that non-thermal radiation mechanisms involving relativistic particles play a crucial role. 

Recently, TeV emission has been detected from GRB afterglows using the High Energy Stereoscopic System (H.E.S.S.) and Major Atmospheric Gamma Imaging Cherenkov (MAGIC) telescopes. Observations reveal the complex nature of the afterglow, which does not follow expected relations. In the case of GRB~190114C, the model required the shock microphysical parameters to evolve with time to explain the afterglow evolution \citep{Misra2021}. Detection of VHE emission from some GRBs, extending into the TeV range, presents novel challenges for GRB models and particle acceleration concepts. Existing afterglow models cannot explain the production of TeV photons. Proposed scenarios for the VHE emission include synchrotron self-Compton, proton synchrotron radiation, or the decay of secondary particles from photohadronic interactions \citep{2001ApJ...559..110Z, 2015MNRAS.454.1073B, PhysRevLett.78.4328, 1998APS..APR..R806B}. It remains unclear if these mechanisms can fully explain VHE-detected GRBs and if they possess exceptionally distinct properties. Comparing the environments and cluster structure of GRBs with and without VHE emission could determine what makes these VHE bursts unique. 

In this work, we study whether GRBs that exhibit VHE emission have distinguishing characteristics by probing their environments and examining similarities between their light curves using machine learning techniques.

\section{Are VHE-detected GRBs similar?}

Thus far, VHE emission has been reported in six GRBs \citep{Fraija2019, Magic2019, HESS2021, Blanch2020a, Blanch2020b, Dzhappuev2022}. The redshift {\it z}, the $T_{90}$ duration in sec, isotropic equivalent energy ($E_{iso}$) in erg, the maximum energy of the photon detected, and the facility which made the detection are listed in Table 1. Understanding the physical mechanisms that generate VHE emission in these sources can provide valuable insights into the phenomenon of this emission from GRBs in general. A key question is whether GRBs that exhibit emission at such extreme energies share characteristics that distinguish them from the overall GRB population.

To investigate this, we studied and compared the properties and environments of the VHE-detected GRBs summarised in Table 1. Further, machine learning algorithms can examine the fine structures in the GRB light curves and cluster them together based on similarities and dissimilarities between them. We analysed the locations of these GRBs in two-dimensional embeddings created using machine learning algorithms applied to {\it Swift}-BAT catalog data\footnote{\url{https://swift.gsfc.nasa.gov/results/batgrbcat/}} \citep{Lien_2016}. The locations of the GRBs in these embeddings allow us to determine if GRBs with VHE emission tend to lie within the same cluster. Any preferential clustering would suggest common underlying factors that can enable their VHE emission \citep{Dimple_2023}.

\begin{table*}
\centering
\begin{minipage}{150mm}
\renewcommand{\arraystretch}{1.5}
\caption{Properties of the TeV detected GRBs}
\end{minipage}
\label{tab:vhe_grb_properties}
\scriptsize{
 \begin{tabular}{|c| c| c| c| c| c| c|} 
 \hline
GRB  & redshift & $T_{90}$ & $E_{\rm iso}$ & Maximum photon & TeV detection  & References \\
  &  & & & energy & facility &  \\

          & (z)         & (s)      & (erg)         & (TeV)       & & \\
\hline
GRB 180720B &  0.654      &  $51.1 \pm 3.0$   &  $6.82_{-0.22}^{+0.24} \times 10^{53}$  &     0.44   &   H.E.S.S. & \cite{Vreeswijk2018,Fraija2019} \\
GRB 190114C &  0.425      &   $\sim$ 116       &   $3.5 \pm 0.1 \times 10^{53}$            &     1        &  MAGIC & \cite{Magic2019}  \\
GRB 190829A & 0.078       &   $57 \pm 3$       &     $\sim 2.0 \times 10^{50}$          &    3.3         &  H.E.S.S. & \cite{HESS2021}\\
GRB 201015A &  0.426      &   $9.78 \pm 3.47$   &     $\sim 3.86 \times 10^{51}$          &      -       &  MAGIC  & \cite{Blanch2020a}\\
GRB 201216C &   1.1     &    $29.95 \pm 0.57$      &   $\sim 6.32 \times 10^{53}$            &    -         &   MAGIC & \cite{Blanch2020b} \\
GRB 221009A$^\dagger$ &   0.151     &   $1068.40 \pm 13.38$       &   $\sim 1.2 \times 10^{55}$            &     251        &   Carpet-2 & \cite{Dzhappuev2022} \\         
\hline
\end{tabular}
\noindent
\newline
\noindent
$^\dagger$ Also detected by MAGIC
}
\end{table*}

\subsection{Environments}
\begin{figure}[!h]
    \centering
    \includegraphics[scale=0.4]{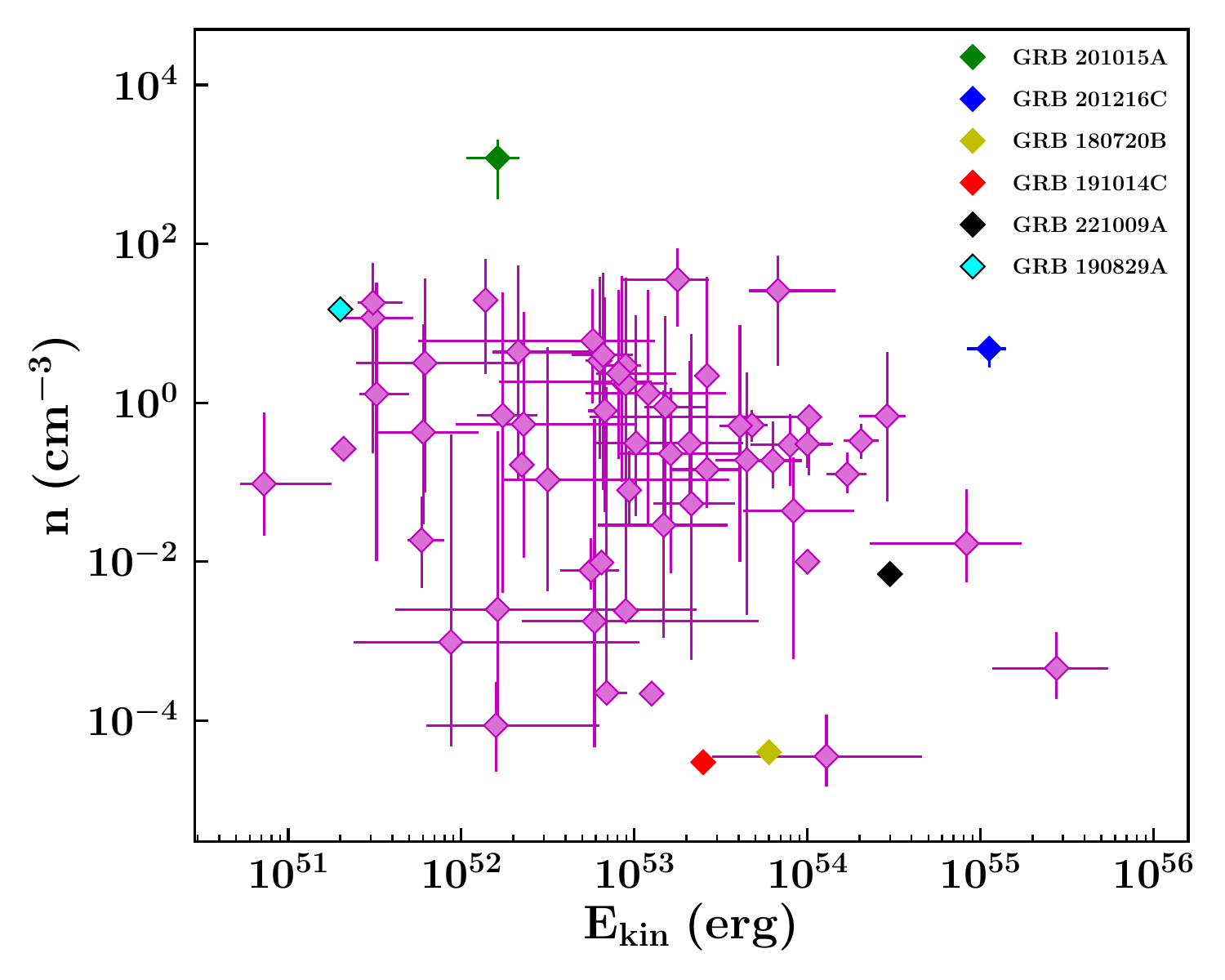}
    \bigskip
    \begin{minipage}{12cm}
    \caption{The relation between the number density and kinetic energy of GRBs is shown in this plot, where the VHE-detected GRBs are indicated with different colors. These GRBs are seen to follow the general trend along with the larger GRB population.}
    \label{fig:VHE_n}
    \end{minipage}
\end{figure}

The densities of the circumburst medium in GRB environments can vary greatly in the VHE regime. GRBs~180720B, 190114C and 221009A have circumburst medium densities around or below 0.1 cm$^{-3}$ \citep{Guarini_2023}. Tentative interpretation of radio, optical, and X-ray data suggest even lower circumburst medium densities of approximately 0.1 cm$^{-3}$. In contrast, GRB~201015A has a mild-relativistic jet surrounded by a very dense medium (n=1202.3 cm$^{-3}$, ~ \citealt{Zhang_2023}), while GRB~201216C has an ultra-relativistic jet surrounded by a moderately dense medium (n=5 cm$^{-3}$, ~ \citealt{Zhang_2023}). GRB~190829A also has {\bf a} moderate environmental density  (n=15 cm$^{-3}$, ~\citealt{Zhang_2021}). Figure \ref{fig:VHE_n} shows the relation between the number density and kinetic energy of GRBs with the VHE detected GRBs highlighted with different colors. We notice that these GRBs follow the general trend of the GRB population, do not exhibit any peculiar behavior, and show no preference for a specific environment. This suggests that VHE GRBs behave similarly to other GRBs regarding their number density and kinetic energy relationship. These findings emphasise the need for further observations and analysis to comprehend the properties of their environment and the implications for understanding these captivating astrophysical phenomena.

\subsection{Locations in two-dimensional embeddings}
Studying populations of GRBs detected by an instrument and analysing their corresponding prompt emission light curves may help identify any structures within GRB populations. This could provide clues on different GRB progenitor classes. Analyzing large numbers of GRB light curves makes the analysis difficult due to high dimensionality. Machine learning algorithms like t-distributed Stochastic Neighbor Embedding \citep[tSNE;][]{JMLR:v9:vandermaaten08a, JMLR:v15:vandermaaten14a} and Uniform Manifold Approximation and Projection \citep[UMAP;][]{McInnes2018} with Principal Component Analysis \citep[PCA;][]{Hotelling1933} initialisation can be used to reduce the dimensionality of the data as well as to visualise the local and global structures within the data. tSNE is a nonlinear dimensionality reduction method that visualises data by embedding high-dimensional neighborhoods stochastically. It minimises the differences between probability distributions in high and lower-dimensional spaces using Kullback-Leibler divergence. On the other hand, UMAP utilises concepts from topology to construct a high-dimensional representation of the data using a similarity matrix. It then leverages these topological relationships to find a low-dimensional projection that preserves distances between points, effectively capturing both local and global structures in the data.

\begin{figure}
    \centering
    \includegraphics[scale=0.2]{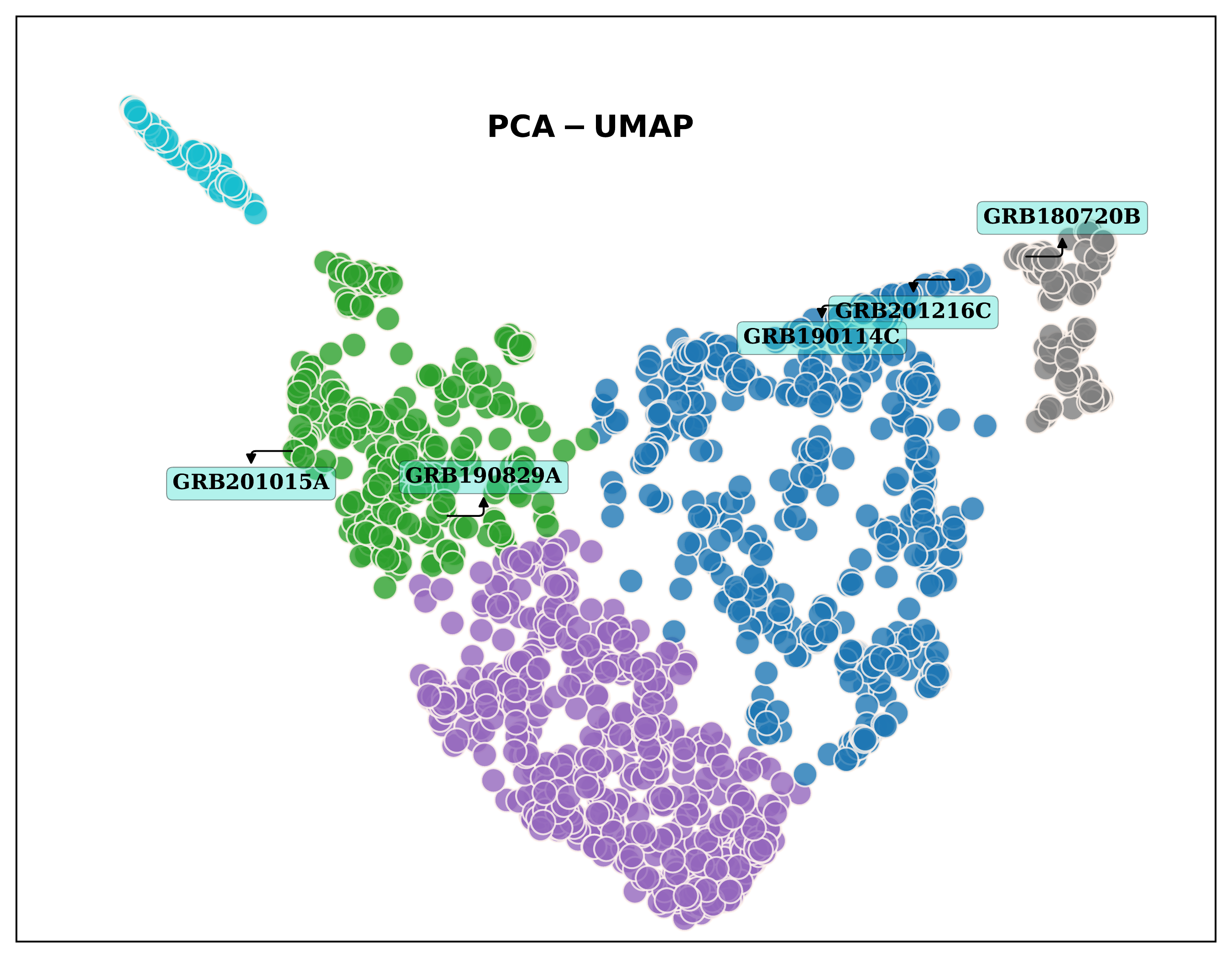}
    \bigskip
    \begin{minipage}{12cm}
    \caption{The location of VHE-detected GRBs in two-dimensional embedding obtained using PCA-UMAP over the {\it Swift}-BAT sample. The different colors indicate the clusters identified by {\sw AutoGMM}.}
    \label{fig:PCA-tSNE/UMAP}
    \end{minipage}
\end{figure}

We employ both tSNE and UMAP with PCA initialisation (PCA-tSNE/PCA-UMAP) to cluster GRB prompt emission light curves. For this, we use the light curves from the Swift-BAT catalog 2022 available in four energy bins (15-25, 25-50, 50-100, and 100-350 keV) with a temporal resolution of 64 ms. This catalog includes light curves of 1525 GRBs, detected between December 17, 2004, and July 15, 2022. The catalog for 2022 consists of light curves of 1525 GRBs detected between December 17, 2004, and July 15, 2022. The data show variations in burst duration, fluence, and start times that can impact machine learning algorithms. Therefore, we standardised the data, which involves normalising the light curves with fluence, shifting them to the same start time, padding them with zeros to match length, and performing the Discrete-time Fourier Transform (DTFT) to retain time delay information between light curves. We could standardise the data only for 1450 GRBs since fluence information was sometimes missing. Next, we performed PCA on the standardised dataset and estimated the number of PCs, preserving around 99\% of the variance, and utilised those for further dimensionality reduction using tSNE and UMAP algorithms. These algorithms were optimised by tuning their key hyperparameters, such as \sw{perplexity} in tSNE and \sw{n\_neighbors} and \sw{min\_dist} in UMAP. We used a \sw{perplexity=25}, \sw{n\_neighbors=25}, and \sw{min\_dist=0.01}, obtained after several iterations, to generate two-dimensional embeddings. The embeddings were further subjected to the Auto Gaussian Mixture Model (\sw{AutoGMM}, \citealt{Athey2019}) algorithm to identify different clusters present. The detailed description can be found in \citet{Dimple_2023}.

Figure \ref{fig:PCA-tSNE/UMAP} shows the two-dimensional embeddings obtained using PCA-UMAP with the locations of VHE-detected GRBs. Similar results are found with PCA-tSNE. The GRBs are located all over the map, giving indications that these GRBs do not cluster in a particular subgroup and do not bear similarities with each other. However, their spread in location all over the map hints that these could be distributed amongst the GRB population, with the only difference being their detection in very high energy regimes. 

\section{Summary}
\label{summary}
In this work, we investigated the properties and environments of GRBs that exhibit VHE emission, a phenomenon that challenges current GRB models and particle acceleration concepts. We have compared the number density and kinetic energy of these GRBs with the overall GRB population and found no significant difference or any peculiar behavior. We have applied machine learning techniques tSNE and UMAP with PCA initialisation (PCA-tSNE/PCA-UMAP) to cluster GRB prompt emission light curves. We created two-dimensional embeddings of GRB prompt emission light curves from the {\it Swift}-BAT catalog and located the VHE-detected GRBs on these maps. We have found that these GRBs are distributed across the map and do not cluster in any particular subgroup, indicating that they do not share similarities among themselves or with any specific progenitor class. Our findings suggest that VHE-detected GRBs behave similarly to other GRBs in terms of their number density, kinetic energy, and light curve morphology and that any distinctive environmental or physical factors do not influence their VHE emission. A larger sample of VHE-detected GRBs is needed to confirm these results and to explore further the origin and mechanism of their high energy emission extending out to TeV bands. Future detections of VHE in GRBs with sensitive detectors will provide more data to advance our understanding of these GRBs.

\begin{acknowledgments}
The authors thank the referee for providing constructive comments on the manuscript. The authors thank Prof. K. G. Arun and Prof. L. Resmi for their useful discussions.
\end{acknowledgments}

\begin{furtherinformation}

\begin{orcids}
\orcid{0000-0003-1637-267X}{Kuntal}{Misra}
\orcid{0000-0001-9868-9042}{Dimple} {}
\orcid{0000-0003-2265-0381}{Ankur}{Ghosh}
\end{orcids}

\begin{authorcontributions}
All authors in this work have made significant contributions.
\end{authorcontributions}

\begin{conflictsofinterest}
The authors declare no conflict of interest.
\end{conflictsofinterest}

\end{furtherinformation}

\bibliographystyle{bullsrsl-en}

\bibliography{extra}

\end{document}